\documentclass[showpacs, oneside, twocolumn, prd, amsmath,amssymb, nofootinbib, 
superscriptaddress]{revtex4-1}
 
\usepackage{cases}
\usepackage{amsmath}
\usepackage{amssymb}
\usepackage{amsfonts}
\usepackage{amssymb}
\usepackage{dcolumn}
\usepackage{bm}
\usepackage{bbm}
\usepackage{graphicx}
\usepackage{xcolor}
\usepackage{array}
\usepackage{subfigure}
\usepackage{hyperref}
\usepackage{wasysym}

\newcommand{\be}{\begin{equation}}
\newcommand{\ee}{\end{equation}}
\newcommand{\ba}{\begin{eqnarray}}
\newcommand{\ea}{\end{eqnarray}}

\newcommand{\gsim}{\mathrel{\hbox{\rlap{\lower.55ex \hbox {$\sim$}}
			\kern-.3em \raise.4ex \hbox{$>$}}}}
\newcommand{\lsim}{\mathrel{\hbox{\rlap{\lower.55ex \hbox {$\sim$}}
			\kern-.3em \raise.4ex \hbox{$<$}}}}

\def\bl#1\el{\begin{align}#1\end{align}}

\hypersetup{colorlinks=true,
	breaklinks=true,
	pdfstartview=Fit,
	linkcolor=blue,
	citecolor=blue,
	urlcolor=blue}

\begin{document}

\title{Data-driven Reconstruction of the Late-time Cosmic Acceleration with $f(T)$ Gravity}

\author{Xin Ren}
\email{rx76@mail.ustc.edu.cn}
\affiliation{Department of Astronomy, School of Physical Sciences, University of Science and Technology of China, Hefei, Anhui 230026, China}
\affiliation{CAS Key Laboratory for Research in Galaxies and Cosmology, University of Science and Technology of China, Hefei, Anhui 230026, China}
\affiliation{School of Astronomy and Space Science, University of Science and Technology of China, Hefei, Anhui 230026, China}

\author{Thomas Hong Tsun Wong}
\email{twht@connect.hku.hk}
\affiliation{Department of Physics, Faculty of Science, The University of Hong Kong, Pokfulam Road, Hong Kong, China}

\author{Yi-Fu Cai\thanks{corresponding author}}
\email{Corresponding author: yifucai@ustc.edu.cn}
\affiliation{Department of Astronomy, School of Physical Sciences, University of Science and Technology of China, Hefei, Anhui 230026, China}
\affiliation{CAS Key Laboratory for Research in Galaxies and Cosmology, University of Science and Technology of China, Hefei, Anhui 230026, China}
\affiliation{School of Astronomy and Space Science, University of Science and Technology of China, Hefei, Anhui 230026, China}

\author{Emmanuel N. Saridakis}
\email{msaridak@phys.uoa.gr}
\affiliation{National Observatory of Athens, Lofos Nymfon, 11852 Athens, Greece}
\affiliation{Department of Astronomy, School of Physical Sciences, University of Science and Technology of China, Hefei, Anhui 230026, China}
\affiliation{CAS Key Laboratory for Research in Galaxies and Cosmology, University of Science and Technology of China, Hefei, Anhui 230026, China}

\begin{abstract}
We use a combination of observational data in order to reconstruct the free function of $f(T)$ gravity in a model-independent manner. Starting from the data-driven determined dark-energy equation-of-state parameter we are able to reconstruct the $f(T)$ form. The obtained function is consistent with the standard $\Lambda$CDM cosmology within $1 \sigma$ confidence level, however the best-fit value experiences oscillatory features. We parametrize it with a sinusoidal function with only one extra parameter comparing to $\Lambda$CDM paradigm, which is a small oscillatory deviation from it, close to the best-fit curve, and inside the $1\sigma$ reconstructed region. Similar oscillatory dark-energy scenarios are known to be in good agreement with observational data, nevertheless this is the first time that such a behavior is proposed for $f(T)$ gravity. Finally, since the reconstruction procedure is completely model-independent, the obtained data-driven reconstructed $f(T)$ form could release the tensions between $\Lambda$CDM estimations and local measurements, such as the $H_0$ and $\sigma_8$ ones. 
\end{abstract}


\maketitle

\section{Introduction} \label{sec:intro}

The concept of dark energy was introduced to explain the acceleration of the expansion of the universe that was discovered in the late 1990s \cite{Riess:1998cb, Perlmutter:1998np}. One of the dark energy candidates is the cosmological constant, leading to the standard cosmological scenario, namely the $\Lambda$CDM paradigm. However, as more and more accurate astronomical data accumulate we could deduce that the standard cosmological model might present some undesirable features. Especially, the tensions that seem to appear in the standard cosmological model parameters derived from different observations, if not resulting from unknown systematics, pose a great challenge to modern cosmology. One of the most significant tensions is the tension of Hubble constant. In particular, the direct measurements by Hubble Space Telescope give $H_0=74.03 \pm 1.42 ~ \mathrm{km} \, \mathrm{s}^{-1} \mathrm{Mpc}^{-1}$ \cite{Riess:2019cxk}, while the Planck 2018 best fit for Cosmic Microwave Background (CMB) data based on $\Lambda$CDM paradigm gives $H_{0}=67.4 \pm 0.5 ~ \mathrm{km} \,\mathrm{s}^{-1} \mathrm{Mpc}^{-1}$ \cite{Akrami:2018vks, Aghanim:2018eyx}. The tension between the two observations has reached 4.4$\sigma$. Another tension that seems to appear is the so-called 
$\sigma_8$ one, which occurs in the measurement of perturbations of large-scale structures and CMB \cite{Delubac:2014aqe, Kazantzidis:2018rnb, Lambiase:2018ows}. In principle, one could follow two main ways to solve these tensions. One is to modify the early evolution of the universe to obtain a relatively small sound horizon at the end of  drag epoch \cite{DiValentino:2020zio}. The other is to modify the late evolution of the universe by replacing the cosmological constant with a dynamic dark energy model such as various scalar-field dark energy models \cite{Tsujikawa:2013fta, 
Caldwell:1999ew,Carroll:2003st, Cai:2009zp} and modified theories of gravity \cite{Sotiriou:2008rp, Cai:2015emx, Capozziello:2011et}. In the same lines, since the physical nature of dark energy remains unknown until today, physicists have put forward many dynamical dark energy theories too and have constructed various specific scenarios.

In order to determine whether the proposed  theories can explain observations, an efficient method is to reconstruct the expression of 
the unknown function that usually appears in a specific model, from   
current cosmological observations \cite{Delubac:2014aqe, Capozziello:2017uam, Dai:2018zwv, Arciniega:2021ffa,Dainotti:2021pqg}. Recent progress on revealing the dark-energy equation-of-state (EoS) parameter as the function of the redshift has paved the way for the reconstruction of the specific dynamical dark energy models \cite{Zhao:2017cud, Wang:2018fng}. Some theoretical models can also give similar EOS parameter evolution \cite{Chimento:2008ws}. Especially, the revealed evolution of EoS displays the crossing of the $-1$ divide. Similarly, other studies with observational constraints on $w$CDM and $w(z)$ have also shown the possibility that $w<-1$ \cite{Capozziello:2018jya, DiValentino:2016hlg, DiValentino:2017zyq, Vagnozzi:2019ezj, Visinelli:2019qqu,Benisty:2020otr}, and in particular that the energy density of dark energy may be negative at high redshifts. Such a behavior might be difficult to be explained using a single scalar field or fluid dark energy models \cite{Cai:2011bs}, and thus inspires us to seek for the modified gravity theory. 

One of the most successful theories of gravitational modification is $f(T)$ gravity \cite{Cai:2015emx}. In contrast to the curvature scalar $R$ of the standard general relativity, the expression of the torsion scalar $T$ in the cosmological background does not contain the time derivative of the Hubble parameter $H$. This feature offers a significant advantage in the reconstruction procedure of $f(T)$ comparing to $f(R)$ gravity, and moreover it has the potential to 
explain the accelerating expansion by using a simple Lagrangian form. $f(T)$ gravity could alleviate both the $H_0$ and $\sigma 8$ tensions from the perspective of effective field theory \cite{Yan:2019gbw, Li:2018ixg}. The perturbation of the early universe and the characteristics of future evolution under the framework of $f(T)$ theory are discussed in \cite{Qiu:2018nle,Bamba:2012vg}. Finally, the constraints on the specific $f(T)$ scenarios due to cosmological observations have been analyzed in detail in 
\cite{Nesseris:2013jea, Basilakos:2018arq, Xu:2018npu, Anagnostopoulos:2019miu, LeviSaid:2020mbb}. 

In this work we are interested in using combined observational data-sets in order to reconstruct the $f(T)$ function. The structure of the manuscript is as follows: We briefly review $f(T)$ gravity in the context of cosmology in Section \ref{sec:ft}. In Section \ref{sec:recons}  we illustrate the observational data sources that we use, and we reconstruct the $f(T)$ function in a model-independent 
way. Then we propose an analytic $f(T)$ form that can describe it. Finally, we summarize our results and we provide a discussion in Section \ref{sec:conclu}.

\section{$f(T)$ gravity and cosmology} \label{sec:ft}

In this section, we provide a brief review on $f(T)$ gravity and its 
application in cosmology. The dynamical variables in $f(T)$ gravity are the tetrad fields $e^{A}{}_{\mu}$, where Greek indices correspond to the spacetime coordinates and Latin indices correspond to the tangent space coordinates. At each point of the spacetime manifold, the tetrad fields  $e^{A}{}_{\mu}$ form an orthonormal basis in the tangent space, which implies that they satisfy the relation 
$g_{\mu \nu}=\eta_{A B} e^{A}{}_{\mu} e^{B}{}_{\nu}$, with $g_{\mu \nu}$ the spacetime metric and where $\eta_{A B}=(1,-1,-1,-1)$ is the tangent-space metric. We mention here that in order to acquire a covariant formulation of $f(T)$ gravity one needs to consider the spin connection too \cite{Krssak:2015oua}. However, for the diagonal tetrad of flat Friedmann-Robertson-Walker metric considered in this work \eqref{eq:metric} lead to vanishing spin connection, and hence we proceed with the form of pure tetrad teleparallel gravity \cite{Krssak:2015oua, Krssak:2018ywd}. 

We consider the Weitzenb$\ddot{\text{o}}$ck connection, defined as 
\begin{equation}
	\hat{\Gamma}^{\lambda}{}_{\mu \nu} \equiv e_{A}{}^{\lambda} \partial_{\nu} 
e^{A}{}_{\mu}=-e^{A}{}_{\mu} \partial_{\nu} e_{A}{}^{\lambda}.
\end{equation}
For this connection, the Riemann curvature vanishes and we only have non-zero torsion, namely
\begin{equation}
T^{\lambda}{}_{\mu \nu} \equiv \hat{\Gamma}^{\lambda}{}_{\nu 
\mu}-\hat{\Gamma}^{\lambda}{}_{\mu \nu}=e_{A}{}^{\lambda}\left(\partial_{\mu} 
e^{A}{}_{\nu}-\partial_{\nu} e^{A}{}_{\mu}\right). 
\end{equation}
Additionally, the torsion scalar is
\begin{equation}
T = S_{\rho }{}^{\mu \nu}  T^{\rho }{}_{\mu \nu},
\end{equation}
where
\begin{equation}
	{S_{\rho}}^{\mu \nu} \equiv \frac{1}{2}\left({{K}^{\mu 
\nu}}_{\rho}+\delta_{\rho}^{\mu} {T^{\alpha \nu}}_{\alpha}-\delta_{\rho}^{\nu} 
{T^{\alpha \mu}}_{\alpha}\right)
\end{equation}
with
\begin{equation}
	{K}^{\rho}{}_{\mu \nu} \equiv 
\frac{1}{2}\left(T_{\mu}{}^{\rho}{}_{\nu}+T_{\nu}{}^{\rho}{}_{\mu}-{T^{\rho}}_{
\mu \nu}\right).
\end{equation}

The simplest theory that one can construct in this framework is the 
teleparallel gravity, whose Lagrangian is the torsion scalar $T$. This theory is equivalent to general relativity at the level of equations of motion, since there exists a transformation relation  between the torsion scalar $T$ and the curvature scalar $R$ \cite{DeAndrade:2000sf, Unzicker:2005in, Aldrovandi:2013wha}. Similarly to $f(R)$ gravity that generalizes the Lagrangian to an arbitrary function of the curvature scalar $R$, one could also generalize the Lagrangian of teleparallel gravity to an arbitrary function of the torsion scalar $T$, which is no longer equivalent to its curvature counterpart. Specifically, the generalized Lagrangian could be written as
\begin{equation}
	S=\int d^{4} x \; e \frac{M_{P}^{2}}{2} [T+f(T)+L_m],
\end{equation}
where $e=det\left(e^{A}{}_{\mu}\right)=\sqrt{-g}$, $M_P$ is the Planck mass and $f(T)$ is the arbitrary function of torsion scalar $T$ (we use units where $c=1$). By varying the above action with respect to the tetrads, we obtain the field equations as 
\begin{align}
\label{tfe}
 & e^{-1} \partial_{\nu}\left(e e_{A}{}^{\rho} S_{\rho}{}^{\mu 
\nu}\right)\left[1+f_{T}\right]-e_{A}{}^{\lambda} T^{\rho}{}_{\nu \lambda} 
S_{\rho}{}^{\nu \mu}\left[1+f_{T}\right]  \nonumber\\ 
&+e_{A}{}^{\rho} S_{\rho}{}^{\mu \nu}\left(\partial_{\nu} T\right) f_{T 
T}+\frac{1}{4} e_{A}{}^{\mu}[T+f(T)] \nonumber\\
&=4 \pi G e_{A}{}^{\rho} T{(m)} _\rho{}^{\mu},
\end{align}
where $f_{T} \equiv \partial f(T) / \partial T$, $f_{T T} \equiv \partial^{2}  f(T) / \partial T^{2}$, and $T{(m)} _\rho{}^{\mu}$ is the matter energy-momentum tensor.

Concerning the background geometry of the universe we consider the flat Friedmann-Robertson-Walker (FRW) metric, which has the form  
\begin{equation}
\label{eq:metric}
d s^{2}=d t^{2}-a^{2}(t) \delta_{i j} d x^{i} d x^{j}, 
\end{equation}
where $a(t)$ is the scale factor. The expression of the torsion scalar under this metric is $T = -6H^2$. Therefore, one advantage of the $f(T)$ gravity is that the torsion scalar $T$ does not contain the time-derivative of the Hubble parameter $H= \dot{a}/a$, which implies that a specific form of $f(T)$ function would connect to the specific 
phenomenological behaviors in an easy way. Inserting the cosmological metric to the field equations \eqref{tfe} we result to the two modified Friedmann equations as \cite{Cai:2015emx}: 
\begin{eqnarray}
\label{fdme}
H^{2} = \frac{8 \pi G}{3} \rho_{m}-\frac{f(T)}{6}+\frac{T f_{T}}{3} \\
\dot{H} = -\frac{4 \pi G\left(\rho_{m}+p_{m}\right)}{1+f_{T}+2 T f_{T T}}. 
\label{fdme22}
\end{eqnarray}
Comparing these two equations with the standard Friedmann equations with the dark energy component, one obtains the effective energy density and pressure of dark energy as
\begin{eqnarray}
\rho_{f(T)}=\frac{M_{P}^{2}}{2}\left[2 T f_{T}-f(T)\right] \label{rhofT1} \\
p_{f(T)}=\frac{M_{P}^{2}}{2}\left[\frac{f(T)-Tf_{T}+2 T^2 f_{T T}}{1+f_{T}+2 T 
f_{T T}}\right]. 
\end{eqnarray}
Finally, the effective EoS parameter of dark energy is defined as
\begin{equation}
\label{wofg}
 w \equiv \frac{p_{f(T)}}{\rho_{f(T)}}=\frac{f(T)-Tf_{T}+2 T^2 f_{T T}}{\left[1+f_{T}+2 T f_{T T}\right]\left[2T f_{T}-f(T)\right]}. 
\end{equation}

\section{Data-driven reconstruction of $f(T)$ function} \label{sec:recons}

In this section we will present a procedure to reconstruct the  involved $f(T)$ function using various datasets. In particular, according to the modified Friedmann equations \eqref{fdme}, \eqref{fdme22}, we can associate a specific $f(T)$ form with the observed data through the Hubble parameter. The Hubble parameter can be obtained from the observational data as a function of the redshift, i.e. $H(z)$ \cite{Cai:2019bdh, Zhang:2016tto, Briffa:2020qli}. By using these data, we can find the relation between the redshift $z$ and $f$, namely $f(z)$. Then we can substitute the expression of $T(z)$ as a function of $z$ into this relation, resulting to the   reconstruction of the specific form of $f(T)$. 

In order to follow the above procedure, we need to first extract the 
expressions for the involved derivatives $f_T$. Since in the variation of the redshift in the observation data $\delta z$ is small, we can make the following approximation:
\begin{eqnarray}
f_{T} \equiv \frac{d f(T)}{d T}=\frac{d f / d z}{d T / d 
z}=\frac{f^{\prime}}{T^{\prime}} \nonumber \\
f^{\prime}(z) \approx \frac{f(z+\Delta z)-f(z)}{\Delta z}. 
\label{approx1}
\end{eqnarray}
Hence, the modified Friedmann equation \eqref{fdme}, assuming dust matter (i.e. $p_m=0$), can be written as
\begin{equation}
    H^{2}(z)=H_{0}^{2}\left[\left(1-\Omega_{M}\right) 
\frac{\rho_{f(T)}(z)}{\rho_{f(T)}(0)}  +\Omega_{M}(1+z)^{3}\right].
\label{Fr1b}
\end{equation}
Now, using Eq.~\eqref{approx1} we can extract the recursive relation between the consecutive redshifts ($z_{i}$ and $z_{i+1}$), namely
\begin{align}
	f\left(z_{i+1}\right)-f\left(z_{i}\right) & =3\left(z_{i+1}-z_{i}\right) 
\frac{T^{\prime}\left(z_{i}\right)}{T\left(z_{i}\right)} \nonumber \\
	& \cdot\left[H^{2}\left(z_{i}\right)-\frac{8\pi 
G}{3}\rho_m(z_i)+\frac{f\left(z_{i}\right)}{6}\right] .
\end{align}
By inserting the expressions of the function $T(z)$, $H(z)$ and $\rho_m(z)$ into the above equation, it can be transformed into
\begin{eqnarray}
\label{re1}
	&&
	\!\!\!\!\!\!\!\!\!\!\!\!\!\!
	f\left(z_{i+1}\right)  =f\left(z_{i}\right) +6\left(z_{i+1}-z_{i}\right) 
\frac{H^{\prime}\left(z_{i}\right)}{H\left(z_{i}\right)}  \nonumber\\
	&&\ \ \ \ \ \ \  \cdot \left[H^{2}\left(z_{i}\right)-H_{0}^{2} \Omega_{m 
0}\left(1+z_{i}\right)^{3}+\frac{f\left(z_{i}\right)}{6}\right] .
\label{fzrecon}
\end{eqnarray}

In summary, we deduce that we could reconstruct the evolution of $f(z)$ in $f(T)$ cosmology, using the $H(z)$ data. Specifically, 
if we have the values of $H$ and $f$ at a given redshift $z_i$, the value of $f$ at the next redshift $z_i$ would be totally determined. Finally, concerning the initial conditions, they can be determined by using the observational values at $z=0$.

\subsection{Numerical Reconstruction}

\begin{figure}[ht]
\includegraphics[width=3.3in]{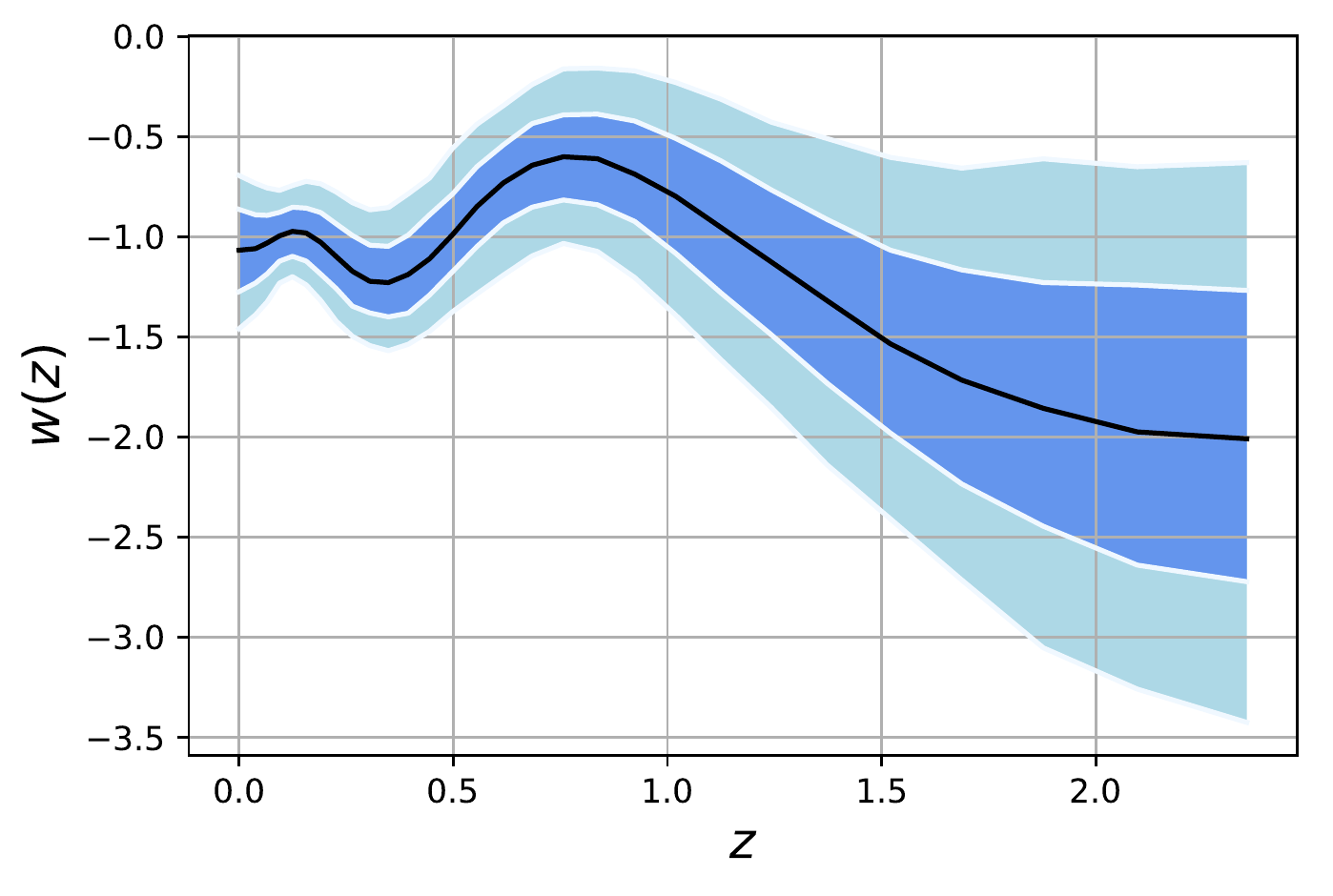}
\caption{{\it{The data-driven reconstructed $w(z)$ of \cite{Zhao:2017cud}. The dark and light blue correspond to $1\sigma$ and $2\sigma$ confidence levels respectively, while the black curve denotes the best-fit value. }}
\label{fig:wz}}
\end{figure}

In this subsection we proceed to the specific application of the above procedure. Observing \eqref{Fr1b} we deduce that in order to calculate the evolution of $f$ we need to insert the values of $w(z)$ reconstructed by the data. This $w(z)$ was reconstructed in \cite{Zhao:2017cud} through a combination of observational data called ALL16, where a Bayesian, non-parametric procedure using the Monte Carlo Markov Chain method was performed. These data-sets ALL16 include the Planck 2015 \cite{Ade:2015xua}, the JLA supernovae \cite{Betoule:2014frx}, the 6dFRS \cite{Beutler:2011hx} and SDSS main galaxy sample BAO measurements \cite{Ross:2014qpa}, the WiggleZ galaxy power spectra \cite{Parkinson:2012vd}, weak lensing from CFHTLenS \cite{Heymans:2013fya}, local measurements of Cepheids \cite{Riess:2016jrr}, $H(z)$ measurements \cite{Moresco:2016mzx},  
BAO and RSD measurements \cite{Zhao:2016das} and Ly$\alpha$ BAO measurements \cite{Delubac:2014aqe}. We are interested in the reconstruction results of the first 29 bins corresponding to redshift $z$ between $0$ and $2.3$ shown in Fig.~\ref{fig:wz} noted as $w_{ALL16}$.

Hence, we can now plug the value of $w(z)$ to each bin and use the modified Friedmann equation \eqref{Fr1b} to solve for the effective energy density and the Hubble parameter. The high-redshift bins solution are depended on the low redshift solution  through \cite{Huterer:2002hy}:
\begin{equation}
 H^{2}(z) = H_{0}^{2} \left[\left(1-\Omega_{M}\right) \frac{\rho_{f(T)}(z)}{\rho_{f(T)}(0)} + \Omega_{M}(1+z)^{3}\right],
\end{equation}
where the $\rho_{f(T)}(z)$ can be represented in terms of $w_j$, for $z$ in bin $j$, as
\begin{eqnarray}
&&
\!\!\!\!\!\!\!\!\!
\rho_{f(T)}(z)=  \rho_{f(T)}\bigg|_{z=0}\left(\frac{1+z}{1+z_{j}-\Delta z_{j} 
/ 2}\right)^{3\left(1+w_{j}\right)} \nonumber \\
&&\ \ \ \    \ \  \ \ \   \   \times\prod_{i=1}^{j-1}\left(\frac{1+z_{i}+\Delta 
z_{i} / 2}{1+z_{i}-\Delta z_{i} / 2}\right)^{3\left(1+w_{i}\right)}.
\end{eqnarray}
Additionally, a set of $H(z)$ and $H'(z)$ can be solved for each sample through the Friedmann equation \eqref{Fr1b}, and then a set of reconstructed $f (z)$ can be obtained using equation \eqref{re1}.
Generating the $w(z)$ samples repeatedly with $w(z)$ mean data and covariance matrix between the different bins, we can obtain the corresponding distribution of $H(z)$, $H'(z)$ and $f(z)$. Then we can acquire the sample distribution range of $1\sigma$ and $2\sigma$ confidence level, as well as the best-fit (mean).

 Since the dark-energy EoS is constant inside each bin, the entire $w(z)$ is not continuous.  If $H'(z)$ is directly solved from Friedmann equation, perfect continuity cannot be guaranteed for $H'(z)$ under the condition that $H(z)$ solved in different bin is continuous. This discontinuity has a great impact on the smoothness of $f_{T}$ and $f_{T T}$. Since our starting point $w(z)$ is correlated with $f_{T}$ and $f_{T T}$ \ref{wofg}, it will lead to non-negligible deviation of the reconstruction results. In order to guarantee the continuity of $H'(z)$, the approach of Gaussian process is considered.

The Gaussian process is a stochastic procedure in order to obtain a collection of random variables, namely to acquire a reconstruction function directly from the known data \cite{Seikel:2012uu}. Such processes can get a set of random variables in which any finite number of variables is subject to a joint normal distribution. The data determine the covariance (kernel) function through training the hyperparameters by maximizing the likelihood function, and then one can obtain the joint normal distribution over functions without assuming any specific model. Gaussian processes are fully defined by their mathematical expectations and kernel functions. Different kernel functions of Gaussian processes would restrict the parameter space and affect the results to varying degrees \cite{Colgain:2021ngq}. We use the squared exponential function, which is the most general form of covariance function, as the kernel function to acquire the $H(z)$ and $H'(z)$, namely
\begin{equation}
 k\left(x, x^{\prime}\right)=\sigma_{f}^{2} e^{-\frac{\left(x-x^{\prime}\right)^{2}}{2 l^{2}}},
\end{equation}
where the $\sigma_{f}$ and $l$ are the hyperparameters. The expectation and kernel functions can be obtained from known data. Hence, applying the Gaussian Processes in Python (GAPP)  we can reconstruct the evolution of functions and their derivatives from the given data points, which has been used extensively in cosmology \cite{Seikel:2013fda,Cai:2015zoa,Wang:2017jdm,Elizalde:2018dvw,Aljaf:2020eqh,Holsclaw:2010sk,Benisty:2020kdt}. Due to the uncertainty of high redshift, we reconstruct the parameters to the redshift range between 0 to 2. In particular, we use GAPP to reconstruct the $H(z)$ and $H'(z)$ up to $z=2$ from the $H(z)$ obtained from the solution of the modified Friedmann equation. This method can avoid the discontinuity among different bins and improve the continuity of $H(z)$ and $H'(z)$.

\begin{figure}[ht!]
\includegraphics[width=3.3in]{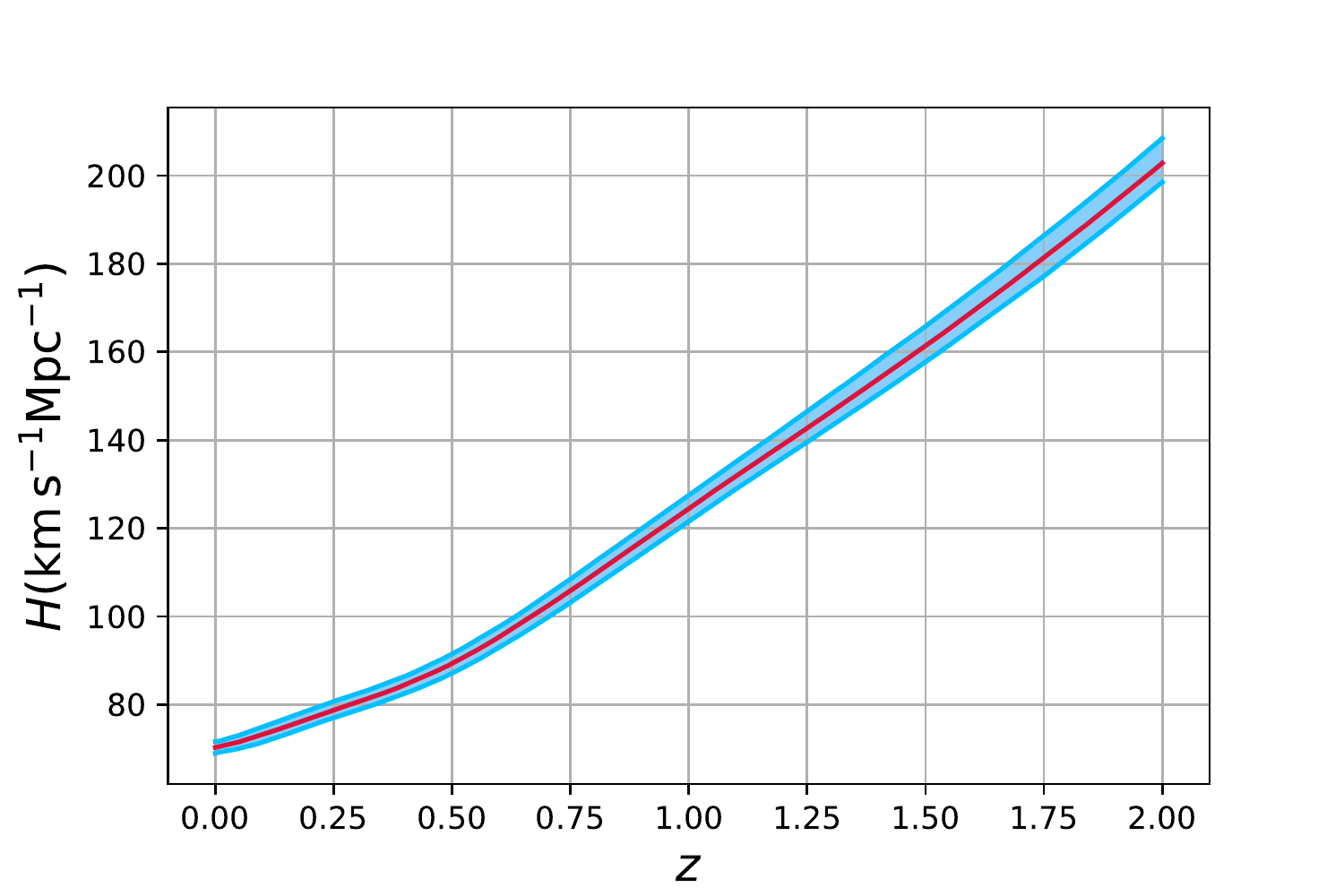}    
\includegraphics[width=3.3in]{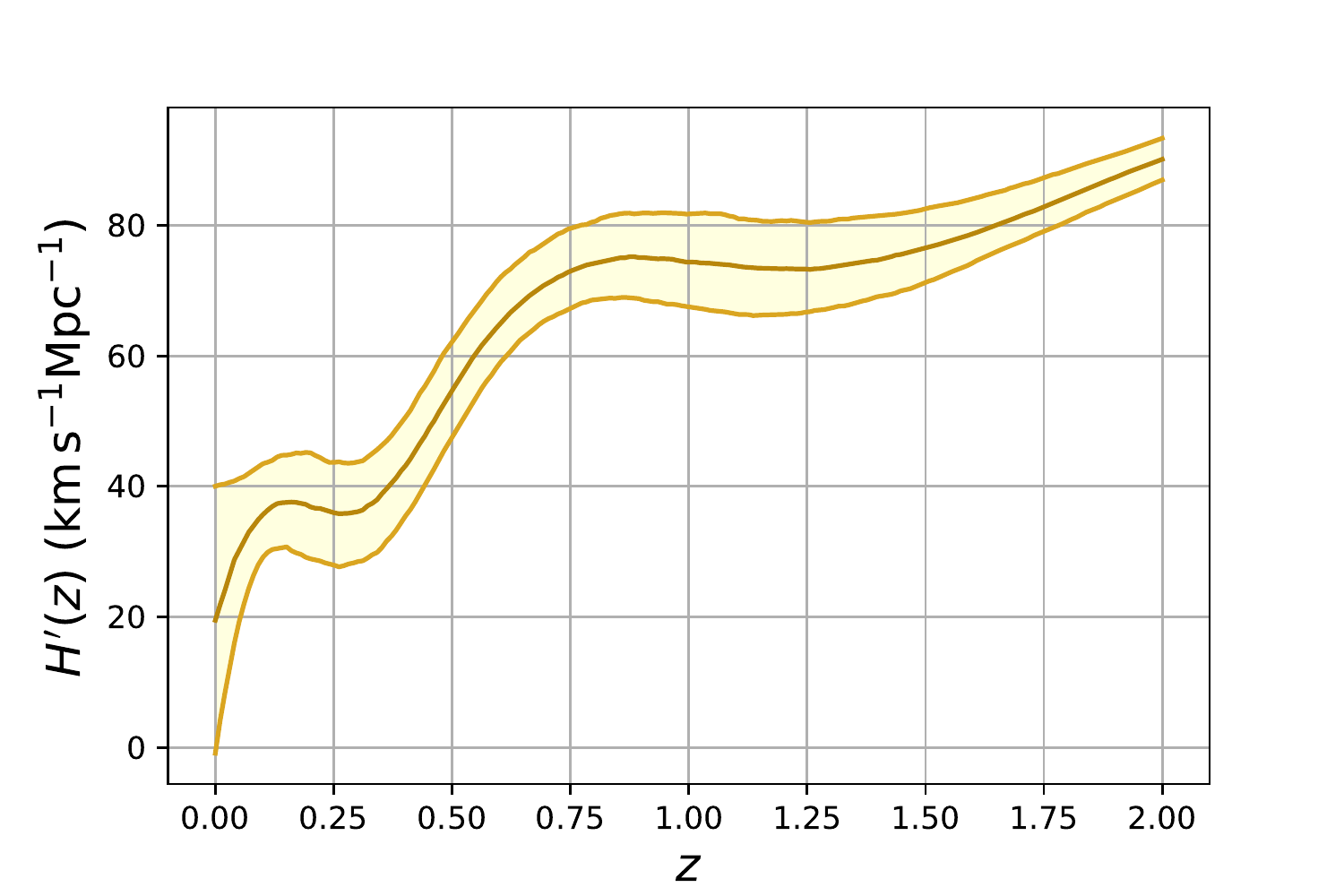} 
\caption{{\it{The reconstructed behavior of $H(z)$ (upper graph) and 
$H'(z)$ (lower graph), arising from the  data-driven reconstructed $w(z)$ of Fig. \ref{fig:wz}, with the present-day values $H_0=70.2 \pm 1.3 \mathrm{km} \, \mathrm{s}^{-1} \mathrm{Mpc}^{-1}$ and $\Omega_{m0}=0.289$. In both graphs, the dark curves denote the best fit, while the shaded area marks the allowed region at $1\sigma$ confidence level. }}
\label{fig:hdh}}
\end{figure}

The best-fit curve, as well as the $1~\sigma$ range, of $H(z)$ and $H'(z)$ from GAPP approach are shown in Fig. \ref{fig:hdh}. Here we choose the present-day values $H_0=70.2 \pm 1.3 ~ \mathrm{km} \, \mathrm{s}^{-1} \mathrm{Mpc}^{-1}$ and $\Omega_{m0}=0.289$ \cite{Zhao:2017cud} as boundary condition to reconstruct the evolution history. Furthermore, the $H'(0)$ can be obtained from the sample of $w(0)$ and $H_0$, and the equation \eqref{fdme}.

\begin{figure}[ht!]
\includegraphics[width=3.3in]{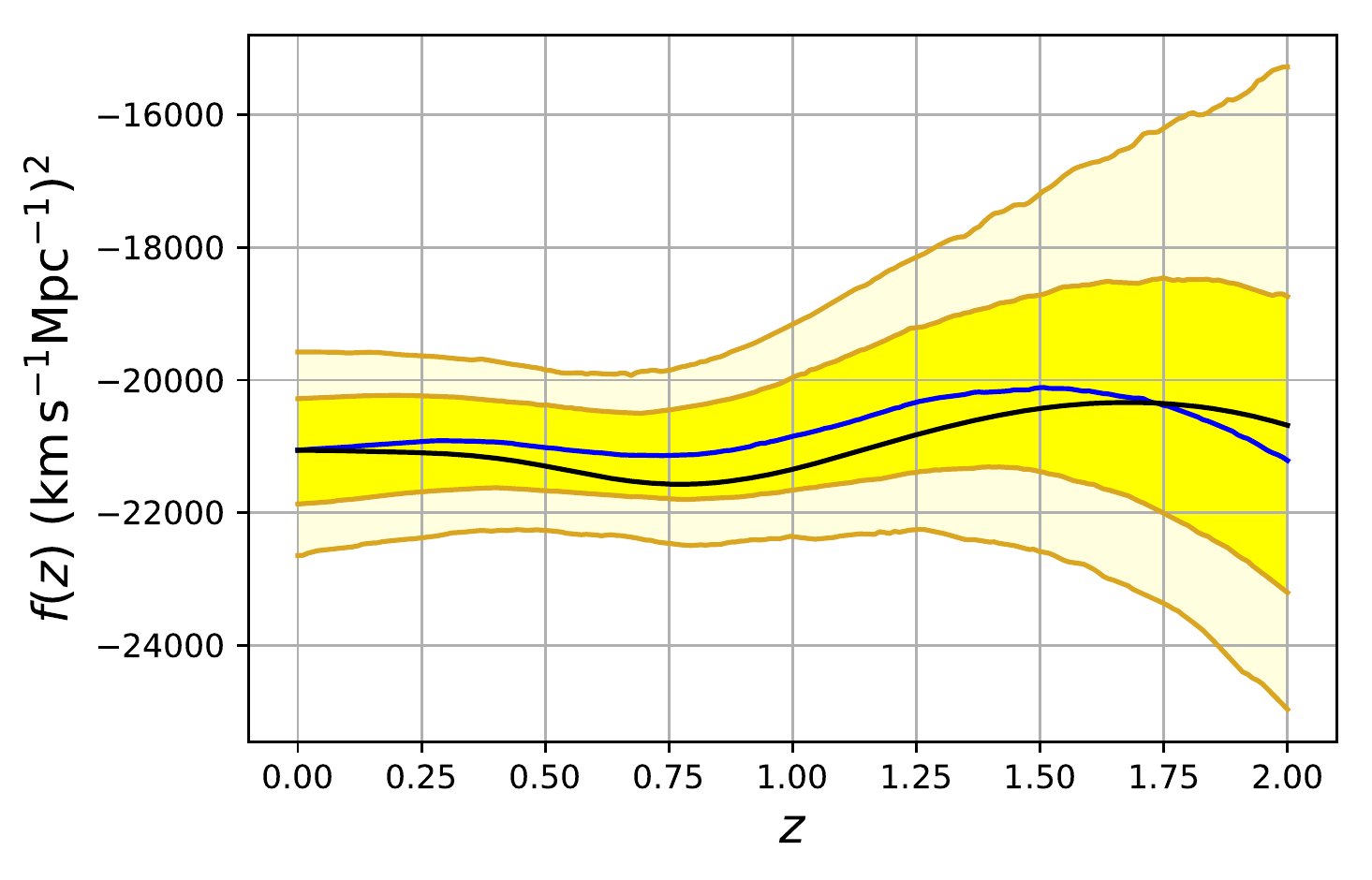}     
\caption{{\it{The reconstructed behavior for $f(z)$ as a function of redshift $z$, arising from the  data-driven reconstructed $H(z)$ and  $H'(z)$ of Fig. \ref{fig:hdh}. The yellow and light yellow regions mark the $1\sigma$ and $2\sigma$ confidence level respectively, the blue curve represents the reconstructed mean values, and the black curve arises using the best-fit curve of $w(z)$ of Fig.~\ref{fig:wz}.}}
\label{fig:g70}}
\end{figure}

\begin{figure}[ht!]
\centering
\includegraphics[width=3.3in]{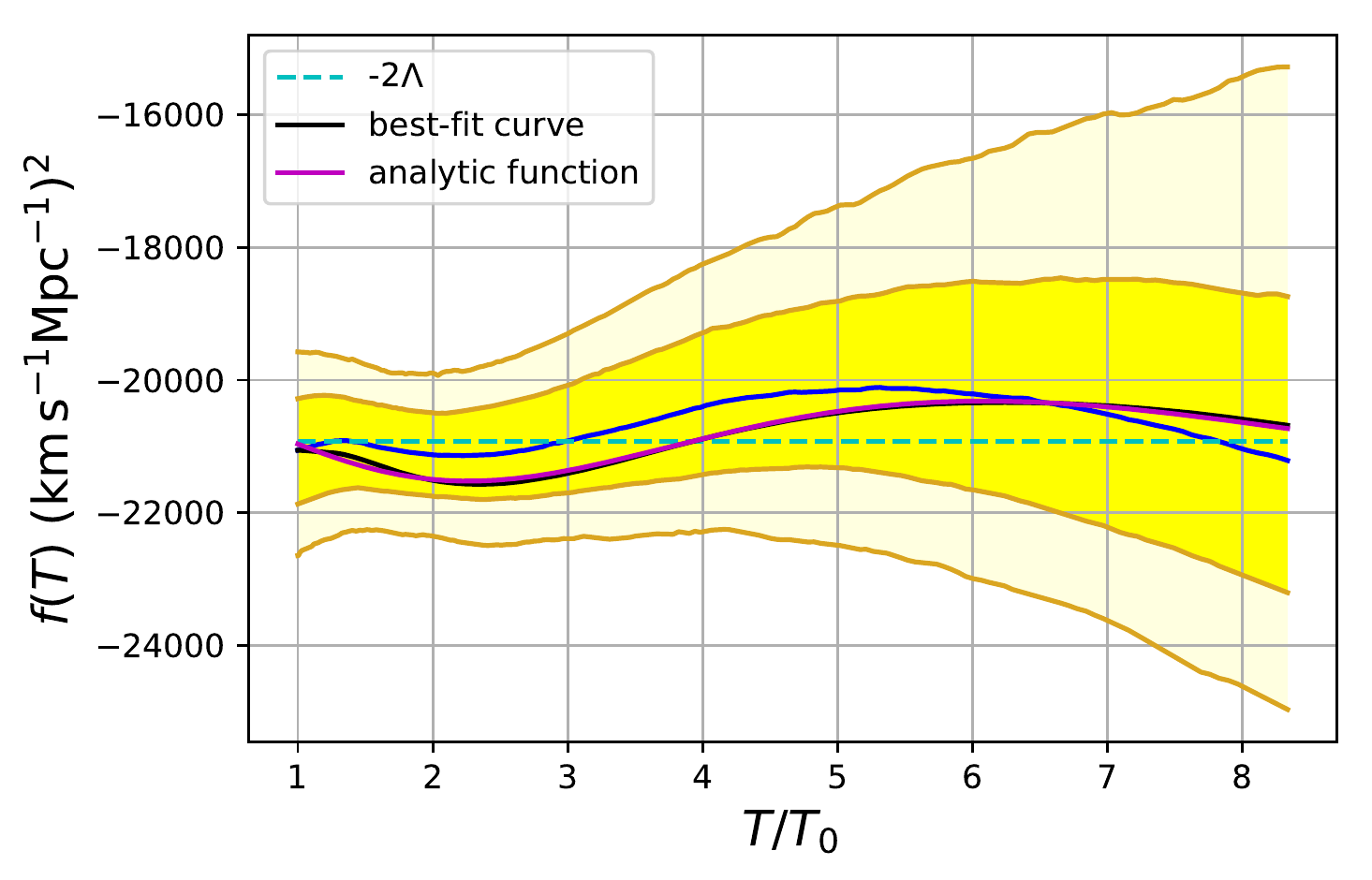}
\caption{\it{The reconstructed $f(T)$ form, arising from the  data-driven reconstructed $f(z)$ of Fig.~\ref{fig:g70}. The yellow and light yellow regions mark the $1\sigma$ and $2\sigma$ confidence level respectively, the blue curve represents the reconstructed mean values, and the black curve arises using the best-fit curve of $f(z)$  
of Fig.~\ref{fig:g70}. Finally, the magenta curve is the analytical 
function given in Eq.~\eqref{eq:fan1}.}}
\label{fig:analy}
\end{figure}

Having reconstructed $H(z)$ and $H'(z)$, we can now proceed to the reconstruction of $f(z)$ distribution using relation \eqref{fzrecon}. In Fig.~\ref{fig:g70} we present the corresponding best-fit curve, as well as the $1\sigma$ and $2\sigma$ regions, for the reconstructed $f(z)$. Now, as mentioned above, knowing $f(z)$ and using the relation between the torsion scalar and the Hubble parameter, namely $T=-6H(z)^2$, it is trivial to convert $f(z)$ to $f(T)$. Hence, in Fig.~\ref{fig:analy} we present the reconstructed $f(T)$ as a function of $T$, where we mention that the units of both $T$ and $f(T)$ are  
$(\mathrm{km} \, \mathrm{s}^{-1} \mathrm{Mpc}^{-1})^2$.
Starting from the data-driven reconstructed $w(z)$ we obtain  the $H(z)$ and $H'(z)$ functions by GAPP (Gaussian Processes in Python), and then the   $f(T)$ function presented in Fig.~\ref{fig:analy}. Note that the Gaussian processes is more sensitive to the overall distribution, and therefore the correlation between each $H(z)$   and 
$H'(z)$ samples will be reduced. This will lead to larger errors in the reconstruction results at high redshift. We mention that we have not assumed any ansatz form for $f(T)$ or any prior for the involved parameters, on the contrary the reconstruction of $f$ is entirely model-independent, and based solely on observational data. This $f(T)$ reconstruction is the main result of the present work.

\subsection{Analytical results}

In this section we proceed by investigating the possible analytic form of the data-driven reconstructed $f(T)$ function. Observing the graph of the reconstructed $f(T)$ function, a first conclusion is that  the constant form $f(T)=-2\Lambda$, which corresponds to the  cosmological constant and thus to $\Lambda$CDM cosmology, lies within the 1$\sigma$ region. This is a cross-check verification of our analysis, and it in agreement with the results of other reconstructed procedures \cite{Cai:2019bdh, Briffa:2020qli}. 

The best-fit for the $f(T)$ function is close to the constant one, nevertheless it presents a slight oscillatory behavior which in turn is capable of describing the oscillatory behavior of the dark-energy EoS parameter arising from the simultaneous consideration of various observational data-sets (see Fig.~\ref{fig:wz}). The sinusoidal function is a good choice for characterizing oscillations. In this case, we need at least three parameters to describe the amplitude, frequency and phase of the oscillation. Observing the detailed form 
of the best-fit curve of Fig.~\ref{fig:analy}, we conclude that we can fit it very efficiently with a function of the form 
\begin{equation}
 f(T) = \alpha T_0 \sin\left(\frac{\beta}{T/T_0 + \delta} - \gamma \right) - 
2\Lambda,
\label{eq:fan1}
\end{equation}
with $T_0=-6H_0^2$, namely a varying sinusoidal function for the oscillation (the first term in \eqref{eq:fan1}) and the rightmost boundary condition owing to its tight constraint (the second term in  \eqref{eq:fan1}). Note that the parameters $\alpha, \beta, \delta, \gamma$ are dimensionless while $\Lambda$ has the units of $T$. Hence, the above $f(T)$ form is a small oscillatory deviation from the  $\Lambda$CDM cosmology. In particular, the exact confrontation 
of the numerically obtained best-fit curve with the above analytical form gives $\alpha T_0 =604\,(\mathrm{km} \, \mathrm{s}^{-1} \mathrm{Mpc}^{-1})^2, \beta=174, \delta=10.5, \gamma=166, \Lambda=10460\,(\mathrm{km} \, \mathrm{s}^{-1} \mathrm{Mpc}^{-1})^2$, while the maximum deviation of our best-fit empirical formula from the numerical data in $y$-axis is $\approx 122$, which is well within the 1$\sigma$ regime. 

Nevertheless, note that the analytical expression \eqref{eq:fan1} is the one that matches the numerically reconstructed best-fit $f(T)$ form perfectly. One can definitely use an oscillatory function with less free parameters that will still be close to the best-fit curve of Fig.~\ref{fig:analy} and deep inside the $1\sigma$ regime. This could be 
\begin{equation}
 f(T) = \alpha T_0\sin\left(\frac{  T_0}{T   }  \right) - 2\Lambda,
\label{eq:fan2}
\end{equation}
which still is a small oscillatory deviation from $\Lambda$CDM cosmology with only one extra free parameter (for $\alpha=0$ $\Lambda$CDM cosmology is recovered), and thus with significantly improve information criteria values, such as the Akaike Information Criterion (AIC), the Bayesian Information Criterion (BIC), and the
Deviance Information Criterion \cite{Anagnostopoulos:2019miu}.

\begin{figure}[ht]
\includegraphics[width=3.3in]{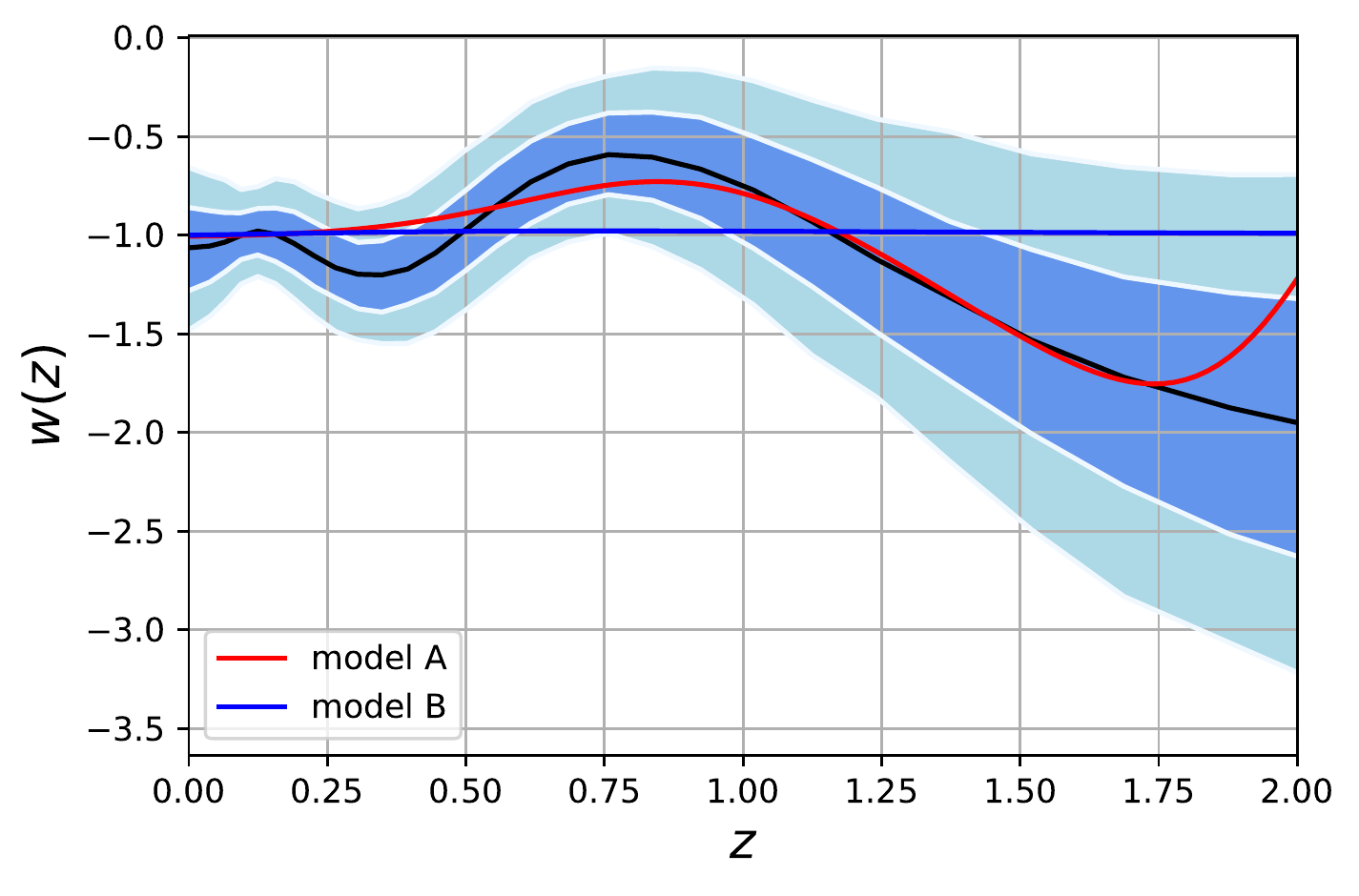}
\caption{{\it{ The distribution of $w_{ALL16}$ and the $w(z)$ derived from model A and model B.}}
\label{fig:wback}}
\end{figure}

As a final cross-check of our analysis, we can now insert the obtained functional form of $f(T)$ into the modified Friedmann equation in order to calculate the resulting dark-energy equation-of-state parameter as a function of redshift. Additionally, we can use the $\chi^2$ to compare the fitting efficiency of   different models. We denote equation \eqref{eq:fan1} as Model A and equation \eqref{eq:fan2} as Model B. The resulting $w(z)$ for the two different models  are presented in Fig.\ref{fig:wback}, on top of the  data-driven reconstructed result $w_ALL16$ of Fig. \ref{fig:wz}. As we can see, Model A  exhibits a clear oscillation behavior, while   Model B is closer to $\Lambda$CDM. Note that there exist some differences between the results of the analysis and the mean of the $w_ALL16$ data. This difference arises mainly from the approximations of the reconstruction process and the limitation to specific function forms.  

Since our reconstruction result focuses on the best-fit curve of $w(z)$ to the redshift $z=2$ (the first 28 bins), the $\chi^2$ of $w_{ALL16}$ can be described as

\begin{equation}
 \chi_{w(z)}^{2}=\sum_{i=1}^{28} \frac{\left(w_{\mathrm{model}, i}-w_{\mathrm{ALL16}, i}\right)^{2}}{\sigma_{w_{i}}^{2}}.
\label{eq:chig}
\end{equation}
Additionally, we define $\Delta\chi^2=\chi^2_{Model}-\chi^2_{\Lambda CDM}$. The results of $\chi^2$ statistic is shown in Tab.\ref{tab:chi}.
As we observe,  Model A and Model B fit the $w_{ALL16}$ data better than the $\Lambda$CDM paradigm. Hence, the obtained models of $f(T)$ gravity can be more efficient in describing the evolution of the Universe .

\begin{table}
\caption{The $\chi^2$ of $w_{ALL16}$ for different models.} \label{tab:chi}
\begin{center}
\begin{tabular}{c|c|c}
\hline\hline
	Model	         & $\chi^2$ & $\Delta\chi^2$ \\   \hline
$\Lambda$CDM & 26.468 & 0  \\	    
Model A & 11.867 & -14.601  \\
Model B & 25.936 & -0.532  \\ 
\hline\hline
\end{tabular}
\end{center}
\end{table}

We mention that similar oscillatory dark-energy scenarios are known to be in good agreement with the observational data \cite{Dodelson:2001fq, Feng:2004ff, Lazkoz:2010gz, Ma:2011nc, Pace:2011kb, Pan:2017zoh}, however up to our knowledge this is the first time that such a behavior is proposed for $f(T)$ modified 
gravity.

\section{Conclusions} \label{sec:conclu}

In this work we used a combination of observational data in order to 
reconstruct the $f(T)$ function of $f(T)$ modified gravity in a 
model-independent manner. Starting from the data-driven reconstructed 
dark-energy EoS parameter of \cite{Zhao:2017cud}, we first reconstructed both $H(z)$ and $H'(z)$ using two methods: the modified Friedmann equations approximation and the Gaussian processes. We found that using Friedmann equations can guarantee the correlation of $H(z)$ and $H'(z)$. This approach can lead to very strong constraints on the reconstruction result. Nevertheless, in particular, since the original $w(z)$ data is divided into bins, in order to ensure that $H(z)$ is continuous between consecutive bins, from the differential equation solution we obtain a slightly discontinuous $H'(z)$. Thus, application of the Gaussian processes provided a continuous reconstructed $H'(z)$. Although the reconstruction results for $f(T)$ form present an increasing uncertainty of the function distribution, which is more significant at higher redshift boundary, the result of reconstruction at mean level is very successful. Comparing the two methods, the Gaussian process can efficiently obtain smooth  $H(z)$ and $H'(z)$, thus leading to very successful reconstruction result of $f(T)$ gravity.

From the reconstructed $H(z)$ and $H'(z)$ we were able to reconstruct $f(z)$ and finally $f(T)$. The obtained data-driven reconstructed function is consistent with the standard $\Lambda$CDM cosmology within  $1 \sigma$ confidence level. However, the best-fit value of the reconstructed model has obvious characteristics of oscillatory evolution. In order to describe these features we parametrized it with an oscillatory, sinusoidal, function, with four free parameters that  indeed leads to a perfect fit. Inspired by this, we then proposed an oscillatory, sinusoidal function with only one extra parameter comparing to $\Lambda$CDM paradigm, which still is a small oscillatory deviation from it, close to the best-fit curve, and definitely inside the $1\sigma$ reconstructed region. Similar oscillatory dark-energy scenarios are known to be in good agreement with observational data, nevertheless this is the first time that such a behavior is proposed for $f(T)$ gravity. Finally, since the proposed model has only one extra free parameter, it is expected to lead to very good information criteria values.

The reconstruction procedure followed above is completely model-independent, especially $\Lambda$CDM-independent, and it is based solely on a collection of intermediate-redshift and low-redshift data-sets. Hence, we expect that the obtained data-driven reconstructed $f(T)$ model could release the tensions between  $\Lambda$CDM estimations and local measurements, such as the $H_0$ and $\sigma_8$ ones. Definitely, a detailed and direct confrontation of the proposed oscillatory function with the data should be performed before we can consider it as a successful modified gravity candidate. Such an analysis lies beyond the scope of the present work and it is left for a future project.

\acknowledgments

We thank Chunlong Li,  Martiros Khurshudyan, Wentao Luo, Yuting Wang and Gongbo Zhao for extensive discussions. This work is supported in part by the National Natural Science Foundation of China (Nos. 11653002, 11961131007, 11722327, 1201101448, 11421303), by the China Association for Science and Technology--Young Elite Scientists Sponsorship Program (2016QNRC001), by the National Youth Talents Program of China, by the Fundamental Research Funds for Central Universities, by the China Scholarship Council Innovation Talent Funds, and by the University of Science and Technology of China Fellowship for International Cooperation. All numerical calculations were operated on the computer clusters {\it LINDA} $\&$ {\it JUDY} in the particle cosmology group at University of Science and Technology of China.

\bibliography{reconstruction}

\end{document}